\documentstyle[11pt]{article}
\title{Nuclear Bag Model and Nuclear Magnetic Moments}
\author{Liang-gang Liu\thanks{Email:stdp05@zsu.edu.cn} \\
Department of Physics, Zhongshan University \\
Guangzhou, 510275, P. R. China}

\date{\today}
\begin{document}
\maketitle
\vspace{2cm}
\begin{abstract}
In 1991, we proposed a model in which nucleus is=20
treated as a spherical symmetric MIT bag and nucleon satisfies
the MIT bag model boundary condition. The model was employed to calculate=20
nuclear magnetic moments. The results are in good agreement with=20
experiment data. Now, we found this model is still interesting and=20
illuminating.
\end{abstract}
\newpage
\section{Introduction}

The MIT bag model, as it's well known, explains the=20
hadronic static properties very well, especially the hadron mass=20
spectra\cite{Chodos}. The most attractive point of the model, as we=20
thought is the linear boundary condition for quarks: $\hat{\bf=20
r}\cdot{\bf j}=3D0$ at the bag surface, where ${\bf j}$ is the quark=20
current. this boundary condition confines the quarks inside the=20
bag absolutely.

Based on the MIT bag model a lot of models have been turned=
 out\cite{Thomas},=20
not only for hadronic static properties but also for dynamical=20
processes. On the other side, the applications of the MIT bag=20
model to nucleus is proposed. The so called quark shell model=20
(QSM)\cite{Petry} treats nucleus with mass number A as 3A quarks moving=20
independently in a big MIT bag. This model can give the correct=20
closed shells and empirical radius rule $R\sim A^{\frac{1}{3}}$.=20
The nuclear magnetic moments are calculated in QSM for some=20
nuclei and orbits\cite{Arima,Kato}. Despite its successes, the QSM is too=20
extreme to believe for ground state nucleus. So in this letter we assume=
 nucleus with a constant density=20
consist of A nucleon which move in a uniformly distributed meson=20
fields produced by $\sigma$ and $\omega$ mesons, and the nucleons are=20
confined in the nucleus by a MIT bag boundary. By calculating the=20
nucleon average binding energy and nuclear magnetic moments we=20
find this naive model can explain experiment data quite well.=20

\section{Nuclear Bag Model}

Assume nucleus forms a spherical symmetric bag with
radius R in which nucleon moves relativistically and satisfies
Dirac equation. Nucleon wave function $\psi_{n\kappa m}$
is given as follows,
\begin{eqnarray}
\psi_{n\kappa m}({\bf r})=3DN_{\kappa}\left [\begin{array}{l}=20
ij_{l}(x_{n\kappa}\frac{r}{R})\Phi_{\kappa m}(\hat{{\bf r}}) \\
-Sgn(\kappa)\beta_{n\kappa}j_{\bar{l}}(x_{n\kappa}\frac{r}{R})\Phi_{-\kappa=
=20
m}(\hat{{\bf r}})
\end{array} \right ],
\end{eqnarray}
here we follow the conventions of Rose\cite{Rose},=20
$\mid\kappa\mid=3D1, 2,\cdots$ and for $\kappa>0$, $j=3Dl-
\frac{1}{2}$, $\bar{l}=3Dl-1$, for$\kappa<0$,$j=3Dl+\frac{1}{2}$,=20
$\bar{l}=3Dl+1$. $N_{\kappa}$ is a normalization constant determined by=20
$\int_{Bag}d\bf{r}\mid\phi_{n\kappa m}(\bf{r})\mid^{2}$=3D1, and=20
given by eq. (11).
\begin{eqnarray}
\beta_{n\kappa}=3D\frac{x_{n\kappa}}
{\tilde{E}(x_{n\kappa})+\tilde{M}_{R}},
\end{eqnarray}
here=20
$\tilde{E}(x_{n\kappa})=3D\sqrt{x_{n\kappa}^{2}+\tilde{M}_{R}^{2}}$,=20
$\tilde{M}_{R}=3D\tilde{M}R$, $\tilde{M}$ is nucleon effective mass=20
in nucleus. We use the empirical rule $R=3Dr_{0}A^{\frac{1}{3}}$,=20
$r_{0}$=3D1.2fm through out the calculation. $x_{n\kappa}$ is the wave=20
number as determined by MIT linear boundary condition=20
$(1+i\hat{{\bf r}}\cdot{\bf \gamma})\psi_{n\kappa m}(r=3DR)$=3D0, that=20
is,
\begin{eqnarray}
j_{l}(x_{n\kappa})=3D-
sgn(\kappa)\beta_{n\kappa}j_{\bar{l}}(x_{n\kappa}).
\end{eqnarray}

The $x_{n\kappa}$, for a number of orbits in the case of=20
\begin{quote}
\begin{description}
\item[Table 1:] The wave number $x_{n\kappa}$ for some orbits=20
($njl$) in the case $\tilde{M}=3DM=3D$938.5MeV, A=3D208.=20
$\tilde{M}=3DM$=3D938.5MeV is given in Table 1,
\end{description}
\end{quote}
\begin{center}
\begin{tabular}{|cc|cc|} \hline
$nlj$    &$x_{x\kappa}$   &$nlj$    &$x_{n\kappa}$  \\  \hline
$0s\frac{1}{2}$  &3.10      &$0h\frac{11}{2}$    &9.21   \\
$0p\frac{3}{2}$  &4.426     &$0h\frac{9}{2}$     &9.23   \\
$0p\frac{1}{2}$  &4.43      &$2s\frac{1}{2}$     &9.29   \\
$0d\frac{5}{2}$  &5.677     &$1f\frac{7}{2}$     &10.26  \\
$0d\frac{3}{2}$  &5.68      &$1f\frac{5}{2}$     &10.28  \\
$1s\frac{1}{2}$  &6.19      &$0i\frac{13}{2}$    &10.35  \\
$0f\frac{7}{2}$  &6.88      &$0i\frac{11}{2}$    &10.38  \\
$0f\frac{5}{2}$  &6.89      &$2p\frac{3}{2}$     &10.746  \\
$1p\frac{3}{2}$  &7.61      &$2p\frac{1}{2}$     &10.75  \\
$1p\frac{1}{2}$  &7.62      &$0j\frac{15}{2}$    &11.47  \\
$0g\frac{9}{2}$  &8.06      &$0j\frac{13}{2}$    &11.51  \\
$0g\frac{7}{2}$  &8.07      &$1g\frac{9}{2}$     &11.53  \\
$1d\frac{5}{2}$  &8.96      &$1g\frac{7}{2}$     &11.55  \\=8A$1d\frac{3}{2}=
$  &8.97      &                    &                                  =20
 \\  \hline
\end{tabular}
\end{center}

\section{Nuclear magnetic moments}

Nuclear magnetic moment is one of the teststone=20
used to examine nuclear models\cite{Arima,Ichii}. For a nucleon in a orbit
(n$\kappa$m), its magnetic moment is defined by
\begin{eqnarray}
{\bf {\mu}}=3D\frac{1}{2}\int_{Bag}{\bf r}\times {\bf j}({\bf r})d{\bf r},
\end{eqnarray}
${\bf j}({\bf r})$is nucleon effective electromagnetic current=20
consist of Dirac current and anomalous nucleon current,
\begin{eqnarray}
{\vec j}=3De\bar{\psi}_{n\kappa m}Q{\vec\gamma}\psi_{n\kappa m}+=20
\frac{e\lambda}{2M}{\vec\bigtriangledown}\times(\bar{\psi}_{n\kappa=20
m}{\vec\sigma}\psi_{n\kappa m}),
\end{eqnarray}
here Q=3D1, $\lambda$=3D1.79 for a proton and Q=3D0, $\lambda$=3D-1.91=20
for a neutron. After some calculations we can split up=20
$\mu=3D\equiv\mu_{z}(m=3Dj)$ into=20
two terms, $\mu=3D\mu_{D}+\mu_{A}$, (in n.m.) where
\begin{eqnarray}
\mu_{D} & =3D &QM_{R}\frac{\beta_{n\kappa}}{x_{n\kappa}}\frac{2j+1}{j+1}
[\bar{l}+\frac{1}{2}+Sgn(\kappa)\frac{x_{n\kappa}\beta_{n\kappa}^{2}}
{x_{n\kappa}+(2l+1)\beta_{n\kappa}Sgn(\kappa)+x_{n\kappa}\beta_{n\kappa}^{2}=
}]=20
\nonumber \\
&   &\cdot[1+\beta_{n\kappa}^{2}(1-\frac{2\beta_{n\kappa}}
{x_{n\kappa}+(2l+1)\beta_{n\kappa}Sgn(\kappa)+
x_{n\kappa}\beta_{n\kappa}^{2}})]^{-1},
\end{eqnarray}
this is consistant with that given in ref.\cite{Arima}. The $\mu_{A}$,=20
due to the anomalous nucleon current contribution, can be=20
expressed as follows,
\begin{eqnarray}
\mu_{A} & =3D &-
\lambda Sgn(\kappa)(\frac{j}{j+1})^{\frac{1}{2}}
N_{\kappa}^{2}[(\frac{j}{j+1})^{\frac{1}{2}Sgn(\kappa)}
\int_{0}^{R}r^{2}drj_{l}^{2}(x_{n\kappa}\frac{r}{R}) \nonumber \\
  &   &+(\frac{j}{j+1})^{-\frac{1}{2}Sgn(\kappa)}
\int_{0}^{R}r^{2}drj_{\bar{l}}^{2}(x_{n\kappa}\frac{r}{R})] +
\frac{1}{2}N_{\kappa}^{2}R^{3}\lambda=20
Sgn(\kappa)(\frac{j}{j+1})^{\frac{1}{2}} \nonumber \\
&   &\cdot[(\frac{j}{j+1})^{\frac{1}{2}
Sgn(\kappa)}j_{l}^{2}(x_{n\kappa})+(\frac{j}{j+1})^{-\frac{1}{2}
Sgn(\kappa)}j_{\bar{l}}^{2}(x_{n\kappa})] \nonumber \\=20
  &   &+\frac{1}{2}N_{\kappa}^{2}R^{3}\lambda Sgn(\kappa)\cdot=20
2j[j](\frac{j}{j+1})^{\frac{1}{2}}(-1)^{l}\left(\begin{array}{lll}
j  &j &1  \\
\frac{1}{2} &-\frac{1}{2} &0=20
\end{array} \right)   \nonumber \\   =20
  &   &\cdot(1+\frac{4}{x_{n\kappa}}\beta_{n\kappa}-
\beta_{n\kappa}^{2})j_{\bar{l}}^{2}(x_{n\kappa}),
\end{eqnarray}
where $[j]=3D\sqrt{2j+1}$ and the last two terms come from the=20
bag surface integration and vanish when $R\rightarrow\infty$,=20
and
\begin{eqnarray}
\frac{1}{2}N_{\kappa}^{2}R^{3} & =3D & { \{j_{\bar{l}}^{2}(x_{n\kappa})
 [1+\beta_{n\kappa}^{2}+(2l+1)\frac{\beta_{n\kappa}}{x_{n\kappa}}
  Sgn(\kappa)]}   \nonumber \\
&   & {\cdot[1+\beta_{n\kappa}^{2}(1-
\frac{2\beta_{n\kappa}}{x_{n\kappa}+(2l+1)\beta_{n\kappa}
Sgn(\kappa)+x_{n\kappa}\beta_{n\kappa}^{2}})]\}}^{-1}.
\end{eqnarray}

The calculated magnetic moments for various odd-A nucleus are=20
shown in Table 2, in this calculation $\frac{\tilde{M}}{M}$=3D0.68=20
is fixed for all nuclei. We can see that there are large=20
\begin{quote}
\begin{description}
\item[table 2:] Magnetic moments of odd-A nuclei. The second and=20
fourth columns are defined in the text and $\mu_{FS}$ is the=20
results of ref.\cite{Nishizaki}, which included the core polarization=20
effect. The last column is the experiment data\cite{Nishizaki}.
\end{description}
\end{quote}
\begin{center}
\begin{tabular} {|c|ccccc|} \hline
Nucleus   &$\mu_{D}$  &$\mu_{A}$  &$\mu$  &$\mu_{FS}$  &Expt \\  \hline
$^{15}N$   &0.483      &-0.688     &-0.205  &-0.149    &-0.283 \\
$^{15}O$   &0.0        &0.734      &0.734    &0.535    &0.719  \\
$^{17}O$   &0.0        &-1.593     &-1.593   &-2.031   &-1.894 \\
$^{17}F$   &3.807      &1.493      &5.301    &4.901    &4.722  \\
$^{39}K$   &1.665      &-1.230     &0.435    &0.448    &0.391  \\
$^{39}Ca$  &0.0        &1.313      &1.313    &0.848    &1.022  \\
$^{41}Ca$  &0.0        &-1.617     &-1.617   &-2.197   &-1.595 \\
$^{41}Sc$  &5.200      &1.516      &6.716    &6.068    &5.430  \\
$^{89}Y$   &0.485      &-0.659     &-0.174   &-0.122   &-0.137 \\
$^{91}Zr$  &0.0        &-5.712     &-5.712  &-2.129    &-1.304 \\
$^{209}Bi$ &5.631      &-1.711     &3.920    &3.731    &4.080  \\
$^{207}Pb$ &0.0        &0.696      &0.696    &0.544    &0.582  \\ \hline
\end{tabular}
\end{center}

cancellations between $\mu_{D}$ and $\mu_{A}$. Second, except=20
$^{91}Z_{r}$. Our results, except $^{91}$Zr,=20
fit experiment data very well.=20
But for isoscalar =8Amagnetic moments the models\cite{Funstahl,Ichii} can=20
fit experiment data=20
better than this model. Comparison of $\mu_{D}$ with that given=20
in ref.\cite{Nishizaki} will be helpful for us to accept this model, two=20
models give about the same value but the calculations in this=20
model are much more simpler.

\section{Conclusions}
=20
Our simple nuclear bag model, as a extension of MIT bag model=20
from hadron to nucleus,=20
can describe nucleon average binding energy, especially=20
nuclear magnetic moments quite well. That means nucleus be=20
thought of as a spherical symmetricbag with sharp boundary=20
is not quite different from its reality. But furtherworksare=20
needed to reproduce single=20
particle energy level correctly, etc.

\end{document}